\documentstyle[a4wide,12pt,epsf]{article}
\begin{document}
% TH FORMAT
\begin{flushright}
\baselineskip=12pt
{SUSX-TH-02-017}\\
%{RHCPP 02-03T}\\
{hep-th/0208ddd}\\
{July  2002}
\end{flushright}
\def\IZ{Z\kern-.5em Z}
\begin{center}
%\vglue 0.5cm
{\LARGE \bf NEW STANDARD-LIKE MODELS 
FROM INTERSECTING D4-BRANES \\}
\vglue 0.35cm
{D.BAILIN$^{\clubsuit}$ \footnote
{D.Bailin@sussex.ac.uk}, G. V. KRANIOTIS$^{\spadesuit}$ \footnote
 {G.Kraniotis@sussex.ac.uk, kraniotis@physik.uni-halle.de} and A. LOVE$^{\diamondsuit}$ \\}
\vglue 0.2cm
	{$\clubsuit$ \it  Centre for Theoretical Physics, University of Sussex\\}
{\it Brighton BN1 9QJ, U.K. \\}
{$\spadesuit$ \it Fachbereich Physik, Martin-Luther-Universit\"{a}t Halle-Wittenberg\\
Friedemann-Bach-Platz 6, D-06099, Halle, Germany\\}
%\vgluw 0.2cm
{$\diamondsuit$ \it  Centre for Particle Physics, Royal Holloway, University of London \\}
{\it Egham,  Surrey TW20-0EX, U.K. }
\baselineskip=12pt

\vglue 2.5cm
ABSTRACT
\end{center}
We construct a new set of intersecting D4-brane models that yield the (non-supersym\-metric) standard model 
up to vector-like matter and, in some cases, extra $U(1)$ factors in the gauge group.
 The models are constrained by the requirement that twisted tadpoles cancel, 
and that the gauge boson coupled to the weak hypercharge $U(1)_Y$ does not get a string-scale mass via a generalised 
Green-Schwarz mechanism.
We find six-stack models 
 that contain all of the Yukawa couplings to the tachyonic Higgs doublets that are needed 
to generate mass terms for the fermions at renormalisable level, but which have charged-singlet scalar tachyons and an unwanted extra 
$U(1)$ gauge symmetry after spontaneous symmetry breaking. There is also a six-stack model without any unwanted 
gauged $U(1)$ symmetries, but which only has the Yukawa couplings to generate masses for the $u$ quarks
 and charged leptons.
A particular eight-stack model is free of charged-singlet tachyons and has gauge coupling 
strengths whose ratios at the string scale  are 
close to those measured at the electroweak scale, consistent with the string scale being at most a few TeV.

%\vglue 0.5cm
{\rightskip=3pc
\leftskip=3pc
\noindent
\baselineskip=20pt

}

\vfill\eject
\setcounter{page}{1}
\pagestyle{plain}
\baselineskip=14pt

The D-brane world offers an attractive, bottom-up route to getting standard-like models from Type II string theory \cite{UTCA}. 
Open strings that begin and end on a stack of $M$ D-branes generate the gauge bosons of the group $U(M)$ living in the world volume of the D-branes.
 So the 
standard approach is to start with one stack of 3 D-branes, another of 2, and $n$ other stacks each having just 1 D-brane, thereby generating 
the gauge group $U(3) \times U(2) \times U(1)^n$. Fermions in bi-fundamental
 representations of the corresponding gauge groups can arise at the intersections of such stacks \cite{BDL}, but to get $D=4$ {\it chiral} fermions 
  the intersecting branes should sit at a singular point in the space transverse to the branes, an orbifold fixed point, for example. In general,
   such configurations yield a non-supersymmetric spectrum, so to avoid the hierarchy problem the
    string scale associated with such models must be no more than a few TeV. Gravitational interactions occur in the bulk ten-dimensional space, and 
    to ensure that the Planck energy has its observed large value, it is necessary that there are large dimensions transverse to the branes \cite{ADD}. 
The D-branes with which we are concerned wrap the 3-space we inhabit and closed 1-, 2- or 3-cycles of a toroidally 
compactified $T^2, \ T^2 \times T^2$ or $T^2 \times T^2 \times T^2$ space. Thus getting the correct Planck scale effectively means that only D4- and D5-brane models are viable, 
since for D6-branes there is no dimension transverse to all of the intersecting branes.  
%In general, such models also have scalar tachyons in 
%their spectrum. These arise because i
In a non-supersymmetric theory the cancellation of the closed-string (twisted) Ramond-Ramond (RR) tadpoles does 
{\it not} ensure the cancellation of the Neveu-Schwarz-Neveu-Schwarz (NSNS) tadpoles.
 There is a resulting instability in the complex structure moduli \cite{BKLO}.
%can either be regarded as a fatal 
%flaw of such theories, or as a stringy realisation of the Higgs model. We prefer the latter view, and leave the stabilisation of the moduli as a problem  
%to be solved later.
 One  way to stabilise some of the (complex structure) moduli is to use an orbifold, rather than a torus, 
for the space wrapped by the D-branes. If the embedding is supersymmetric, then the instabilities are removed. This has been studied \cite{Cvetic},
 using D6-branes, 
 but it has so far proved difficult to get realistic phenomenology consistent with experimental data from such models.

During the past year orientifold models with intersecting D6- and D5-branes
have been constructed that yield 
precisely the fermionic spectrum of the standard model (plus three generations of right-chiral neutrinos) \cite{IMR, CIM}. 
(Other recent work on  intersecting brane models, both supersymmetric 
and non-supersymmetric, and their phenomenological implications may be found in \cite{IBM}.)
The spectrum includes 
 $SU(2)_L$ doublet scalar tachyons that may be regarded as the Higgs doublets that break the electroweak symmetry group, but also, unavoidably,
 colour-triplet and charged singlet tachyons either of which is potentially fatal for the phenomenology. In a previous paper 
\cite{BKL1} 
we studied D4-brane models, transverse to a $\IZ _3$ orbifold,  having three generations of chiral matter that were constrained to have no colour triplet or charged singlet 
scalar tachyons, and which contained the Yukawa couplings to the  Higgs doublets needed to give masses to all quarks and leptons; these models, unavoidably,
also possessed extra, vector-like leptonic matter. The wrapping numbers of the various stacks are constrained by the requirement 
 of RR tadpole cancellation and also by the requirement that the mixed $U(1)_Y \times SU(2)_L^2$ and $U(1)_Y \times SU(3)_c^2$ anomalies cancel.
  In string theory, these latter constraints derive from a generalised Green-Schwarz mechanism that ensures that the gauge bosons 
  associated with all anomalous $U(1)$s acquire string-scale masses \cite{IRU}. In fact the cancellation of anomalies is necessary, 
  but {\it not} sufficient \cite{IMR}, to
  ensure the masslessness of the gauge boson associated with $U(1)_Y$. 
  When the stronger constraints derived from this Green-Schwarz mechanism are applied we find 
  %We have recently realised 
  that in the models constructed in \cite {BKL1} the 
  weak hypercharge  $U(1)_Y$ survives only as a global, {\it not} a local, symmetry\footnote{A similar observation applies also to 
  the models constructed in \cite{AFIRU2} and Kataoka and Shimojo \cite{IBM}.}. This paper, therefore, is the first to construct semi-realistic intersecting D4-brane orbifold models that 
  do not have this defect.
   
As before \cite{BKL1}, and as displayed in Table 1, we start with an array of D4-brane stacks, each wrapping a closed 
1-cycle of $T^2$, and situated
 at a fixed point of the transverse $(T^2 \times T^2)/\IZ_3$ orbifold.  
The generator $\theta$ of the $\IZ_3$ point group is embedded in the stack $a$ of $N_a$ branes as
 $\gamma _{\theta,a} = \alpha ^{p_a}I_{N_a}$, where $\alpha = e^{2\pi i/3}, \ p_a=0,1,2$, and the weak hypercharge $Y$ is given \cite{BKL1} by
 \begin{equation}
 -Y = \frac{1}{3}Q_1 + \frac{1}{2}Q_2 + \sum_{i_1}Q_{i_1} + \sum_{j_1}Q_{j_1}+\sum_{k_1}Q_{k_1}
 \label{Y}
 \end{equation}
 where $Q_a$ is the $U(1)$ charge associated with the stack $a$; $Q_a$ is normalised such that the ${\bf N}_a$ 
 representation of $SU(N_a)$ has $Q_a=+1$. The wrapping numbers $n_a$ and $m_a$ specify the number of 
times the two basis 1-cycles are wrapped by the closed 1-cycle associated with stack $a$. When $n_a$ and $m_a$ are coprime 
a single copy of the gauge group $U(N_a)$ occurs;  if $n_a$ and $m_a$ have a highest common factor $f_a$ there are $f_a$ copies of  $U(N_a)$. 
The wrapping numbers determine the number of intersections $I_{ab}$ of stack $a$ with stack $b$
\begin{equation}
I_{ab} = n_am_b - n_bm_a
\end{equation}
 \begin{table}
%\begin{eqnarray}
\begin{center}
\begin{tabular}{|c|c|c|c|} \hline \hline
Stack $a$ & $N_a$ & $(n_a,m_a)$ & $\gamma_{\theta,a}$ \\
\hline \hline
1 & 3 & $(1,0)$ & $\alpha ^p {\bf I}_3$ \\
2 & 2 & $(n_2,3)$ & $\alpha ^q {\bf I}_2 $\\
$i_1 \in I_1$ & 1 & $(n_{i_1},m_{i_1})$ & $\alpha ^q $\\
$i_2 \in I_2$ & 1 & $(n_{i_2},m_{i_2})$ & $\alpha ^q $\\
$j_1 \in J_1$ & 1 & $(n_{j_1},m_{j_1})$ & $\alpha ^r $\\
$j_2 \in J_2$ & 1 & $(n_{j_2},m_{j_2})$ & $\alpha ^r $\\
$k_1 \in K_1$ & 1 & $(n_{k_1},m_{k_1})$ & $\alpha ^p$ \\
$k_2 \in K_2$ & 1 & $(n_{k_2},m_{k_2})$ & $\alpha ^p$ \\
\hline \hline
\end{tabular}
\end{center}
\caption{Multiplicities, wrapping numbers and Chan-Paton phases for the D4-brane models, 
($p\neq q \neq r \neq p$).}
%\end{eqnarray} 
\end{table}
The cancellation of twisted tadpoles gives the constraints:
\begin{eqnarray}
2n_2+\sum_{i_1}n_{i_1}+\sum_{i_2}n_{i_2}=\sum_{j_1}n_{j_1}+\sum_{j_2}n_{j_2} =3+\sum_{k_1}n_{k_1}+\sum_{k_2}n_{k_2} \label{n}\\
6+\sum_{i_1}m_{i_1}+\sum_{i_2}m_{i_2}= \sum_{j_1}m_{j_1}+\sum_{j_2}m_{j_2}=\sum_{k_1}m_{k_1}+\sum_{k_2}m_{k_2}  \label{m}
\end{eqnarray}
and this is sufficient to ensure cancellation of the cubic non-abelian anomalies. The mixed $U(1)$ anomalies are cancelled by
 a generalised Green-Schwarz mechanism involving the exchange of fields that arise from the dimensional reduction of the twisted two-form RR fields 
 $B_2^{(k)}$ and $C_2^{(k)}$ that 
 live at the orbifold singularity \cite{AFIRU1}. These fields are coupled to the $U(1)_a$ field strength $F_a$ of the stack $a$ by terms in the low energy 
 action of the form 
 \begin{eqnarray}
 n_a \int _{M_4}{\rm Tr}(\gamma _{k,a} \lambda _a)C_2^{(k)} \wedge {\rm Tr}F_a \\
 m_a \int _{M_4}{\rm Tr}(\gamma _{k,a} \lambda _a)B_2^{(k)} \wedge {\rm Tr}F_a  
 \end{eqnarray}
 where $\gamma _{k,a} \equiv \gamma _{\theta,a}^k $ and $\lambda _a$ is the Chan-Paton matrix associated with the $U(1)$ generator. 
 These couplings determine the linear combination of $U(1)$ gauge bosons that acquire string-scale masses via the Green-Schwarz mechanism. 
 For the D4-brane array given in Table 1 the couplings are 
 \begin{eqnarray}
 \left[ \alpha^{pk} \left( 3F_1+\sum_{k_1}n_{k_1}F_{k_1} +\sum_{k_2}n_{k_2}F_{k_2}-\sum_{j_1}n_{j_1}F_{j_1}-\sum_{j_2}n_{j_2}F_{j_2} \right)  \right. \nonumber \\
 \left. + \alpha^{qk} \left( 2n_2F_2+\sum_{i_1}n_{i_1}F_{i_1} +\sum_{i_2}n_{i_2}F_{i_2}-\sum_{j_1}n_{j_1}F_{j_1}-\sum_{j_2}n_{j_2}F_{j_2} \right) \right] \wedge C_2^{(k)} \\
 \left[ \alpha^{pk} \left( \sum_{k_1}m_{k_1}F_{k_1} +\sum_{k_2}m_{k_2}F_{k_2}-\sum_{j_1}m_{j_1}F_{j_1}-\sum_{j_2}m_{j_2}F_{j_2} \right)  \right. \nonumber\\
 \left. +\alpha^{qk} \left( 6F_2+\sum_{i_1}m_{i_1}F_{i_1} +\sum_{i_2}m_{i_2}F_{i_2}-\sum_{j_1}m_{j_1}F_{j_1}-\sum_{j_2}m_{j_2}F_{j_2} \right) \right] \wedge B_2^{(k)} 
  \end {eqnarray}
We require that the $U(1)_Y$ gauge boson associated with the weak hypercharge given in eqn (\ref{Y}) remains massless. Consequently, the corresponding 
field strength must be orthogonal to those that acquire Green-Schwarz masses. Thus we require that the wrapping numbers satisfy the constraints:
\begin{eqnarray}
n_2+ \sum _{i_1}n_{i_1}   =   \sum _{j_1}n_{j_1} = 1+ \sum _{k_1}n_{k_1}  \label{nj1} \\ 
3+  \sum _{i_1}m_{i_1} =  \sum _{j_1}m_{j_1} =  \sum _{k_1}m_{k_1} \label{mj1}
\end{eqnarray}
Combining these with the tadpole cancellation constraints (\ref{n},\ref{m}) gives also
 \begin{eqnarray}
n_2+ \sum _{i_2}n_{i_2}   =  \sum _{j_2}n_{j_2} = 2+ \sum _{k_2}n_{k_2}\label{nj2} \\
3+  \sum _{i_2}m_{i_2} =  \sum _{j_2}m_{j_2} =  \sum _{k_2}m_{k_2} \label{mj2}
\end{eqnarray}

In general, tachyonic scalars arise at intersections between stacks $a$ and $b$ which have the same Chan-Paton factor $p_a=p_b$. Thus,
 Higgs doublets, which are needed to give mass to the fermionic matter, arise at $(2i_1)$ and $(2i_2)$ intersections\footnote{
 In our model, as in all others, the Higgs content is non-minimal; this seems to be a generic feature of models deriving from string 
 theory \cite{GN}.}. 
 The allowed Yukawas satisfy selection rules that derive from a $Z_2$ symmetry associated with each stack of ${\rm D}4$-branes.
 A state associated with a string 
between the $a$th and $b$th stack of ${\rm D}4$ branes is odd under 
the $a$th and $b$th $Z_2$ and even under any other $Z_2$.
 We seek first 
 the most economical model that has $n_G=3$ generations of chiral matter, and
  allows the Yukawa couplings needed to give renormalisable-level masses to all (chiral and vector-like) matter. 
  The general solution of eqns (\ref{nj1},
   \ref{mj1}, \ref{nj2}) and (\ref{mj2})
with at most one stack of each type is given in Table 2.
 It is easy to check that these solutions satisfy eqns (11) and (12) of \cite{BKL1} that ensure the absence of
  mixed $U(1)_Y \times SU(2)_L^2$ and $U(1)_Y \times SU(3)_c^2$ anomalies.
% Note that the solution has the property that
%\begin{equation}
%I_{2i_1}=I_{2j_1} \quad {\rm and} \quad I_{2i_2}=I_{2j_2}
%\label{2i2j}
%\end{equation}
  Stacks $1,2,i_1$ and $i_2$ are needed to generate
 the 3 quark doublets and (tachyonic) Higgs doublets.
 % and it is easy to see that 
 %at least two further stacks are also needed. Obviously models with just $j_1$ and $k_1$ as the additional sectors are inconsistent,
 % as are models with just $j_2$ and $k_2$. The four other six-stack models all have $m_{j_1}=0=m_{j_2}$, so if $n_{j_1}=0$ 
 % then $I_{2j_1}=0$, and then from (\ref{2i2j}) there are no Higgs doublets from the $(2i_1)$ intersection; similarly if 
 % $n_{j_2}=0$, there are no Higgs doublets from the $(2i_2)$ intersection. Since both types of Higgses are needed for the Yukawa interactions,
 %  we conclude that the only viable six-stack model has additional $j_1$ and $j_2$ sectors with $n_{j_1}=1$ and $n_{j_2}=2$. 
 These four stacks are also sufficient to provide 3 copies of $u^c_L$ and $d^c_L$ with mass terms at renormalisable level for all quarks.
  To obtain leptons with renormalisable mass terms a fifth stack, $j_1,j_2,k_1$ or $k_2$, is necessary. However, 
   a sixth stack is required for consistency with eqns (\ref{nj1},
   \ref{mj1}, \ref{nj2}) and (\ref{mj2}). Assuming for the moment that the sixth stack is in a different class from the first five, 
   it is easy to see from Table 2 that there are just four six-stack solutions. These have $j_1,j_2$ or 
   $j_1,k_2$ or $k_1,j_2$ or $k_1,k_2$ stacks besides the four above, and all have $m_{j_1}=0=m_{j_2}$. 
   They have $(n_{j_1},n_{j_2})=(1,2)$, 
   $(1,0)$, $(0,2)$, and $(0,0)$ respectively. Of these only the first has $I_{2i_1} \neq 0$ and 
   $I_{2i_1} \neq 0$ so as to provide Higgs bosons which are able to give masses to all quark and lepton matter consistently with the above selection rules.
 \begin{table}
%\begin{eqnarray}
\begin{center}
\begin{tabular}{|c|c|c|c|} \hline \hline
Stack $a$ & $N_a$ & $(n_a,m_a)$ & $\gamma_{\theta,a}$ \\
\hline \hline
1 & 3 & $(1,0)$ & $\alpha ^p {\bf I}_3$ \\
2 & 2 & $(n_2,3)$ & $\alpha ^q {\bf I}_2 $\\
$i_1$ & 1 & $(n_{j_1}-n_2,m_{j_1}-3)$ & $\alpha ^q $\\
$i_2$ & 1 & $(n_{j_2}-n_2,m_{j_2}-3)$ & $\alpha ^q $\\
$j_1$ & 1 & $(n_{j_1},m_{j_1})$ & $\alpha ^r $\\
$j_2$ & 1 & $(n_{j_2},m_{j_2})$ & $\alpha ^r $\\
$k_1$ & 1 & $(n_{j_1}-1,m_{j_1})$ & $\alpha ^p$ \\
$k_2$ & 1 & $(n_{j_2}-2,m_{j_2})$ & $\alpha ^p$ \\
\hline \hline
\end{tabular}
\end{center}
\caption{Multiplicities, wrapping numbers and Chan-Paton phases for D4-brane models consistent with eqns (\ref{nj1},
   \ref{mj1}, \ref{nj2}, \ref{mj2}). }
%\end{eqnarray} 
\end{table}
As before \cite{BKL1}, the three generations of chiral matter include right-chiral neutrino states, and
 there is additional vector-like leptonic, but not quark, matter. We find
\begin{equation}
12(L+\bar{L})+6(e^c_L +\bar{e}^c_L)+3(\nu^c_L +\bar{\nu}^c_L)
\end{equation}
 In addition there are 3 Higgs doublets at the $(2i_1)$ intersections, and 6 at $(2i_2)$. Less welcome
  are the 3 charged-singlet tachyons that arise at the $(i_1i_2)$ intersections.  In general, intersecting brane models also generate 
 colour-triplet tachyons (at the  $(1k_1)$ and $(1k_2)$ intersections), but they are absent in this six-stack model. As noted below, 
 the ratios of the gauge coupling strengths at the string scale also depend on the wrapping numbers. We shall see that the six-stack 
 model that we have derived gives values for these ratios that are inconsistent with those measured at the electroweak scale. 
 Nevertheless, we saw before \cite{BKL1} that the contributions of the extra vector-like leptons and gonions 
 to the renormalisation group equations in such models are such that these values can easily be accommodated
  consistently with a string scale of no more than 3 TeV.  So the models should not be excluded on these grounds.

 A more serious objection is the presence of charged-singlet tachyons. 
 Even when colour-triplet tachyons arise, strong radiative corrections might be sufficient to render them non-tachyonic \cite{CIM}, but 
 charged-singlet tachyons are more difficult to remove. A possible way round this difficulty could be to make the squared-mass 
 $m_{\phi}^2$ of the colour-singlet tachyons numerically small compared with those $m_{H_{1,2}}$  of the Higgs doublets. In general, the squared-mass of the tachyons 
 at an intersection of stack $a$ with stack $b$ is
 \begin{equation}
 m_{ab}^2= -\frac{m_s^2 \epsilon |I_{ab}|R_2/R_1}{2\pi|n_a-m_aR_2/R_1| |n_a-m_aR_2/R_1|}. 
 \end{equation}
 where $m_s$ is the string scale, $R_1$ and $R_2$ are the radii of the two fundamental 
1-cycles of the torus wrapped by the D4-branes, and  
\begin{equation}
\epsilon \equiv 2|\cos \theta/2| \label{eps}\\
\end{equation}
 with $\theta$ is the angle between the two vectors defining the lattice. 
 The squared-masses $m_{H_{1,2}}^2$ of the Higgs doublets at the $(2i_1)$ and $(2i_2)$ intersections 
  both have the factor $|\delta|$, 
 where
 \begin{equation}
 \delta \equiv n_2-3R_2/R_1 \label{del}
 \end{equation}
 in the denominator arising from the $a=2$ stack. Because charged-singlet tachyons are not at an intersection with the $a=2$ stack, $|\delta|$ does not 
 occur in the denominator for these states. Consequently, by taking $|\delta|$  to be small \cite{AFIRU2}, the 
  mass-squared of the charged-singlet tachyons is numerically small compared with that of each Higgs doublet, and could easily 
  be turned positive by radiative corrections. Unfortunately, $|\delta|$ also occurs in eqn (\ref{32}), and a small value of $|\delta|$ would 
  lead to a value of $\alpha_3(m_s)/\alpha_2(m_s)$ too small to run to empirical values with a modest value of $m_s$. 
  Another possible way around the difficulty without deploying more than  six stacks might be to use the five stacks needed to get quark 
  and lepton generations with renormalisable masses, plus a sixth stack in the same class as one already employed, other than $a=1$ or $a=2$. 
  However, it is not difficult to see using Table 2 that such models are inconsistent with eqns (\ref{nj1},
   \ref{mj1}, \ref{nj2}) and (\ref{mj2}). It remains possible that radiative corrections turn the charged-singlet tachyons non-tachyonic 
   while the Higgs doublet scalars remain tachyonic even for $|\delta| \simeq 3.54$. Whether this is the case or not depends upon the 
   contributions of Kaluza-Klein modes, winding modes and gonions to the renormalisation group coefficients. These can be substantial 
   for a low string scale.
 
 There are further objections to these models regarding the number of surviving $U(1)$ gauged symmetries. 
 As already noted, the gauge group associated with the stack $a$ is $U(N_a)$, provided $n_a$ and $m_a$ are coprime, and if this 
 were the case for all six stacks we should start with  6 $U(1)$s. However, since $(n_{j_2}, m_{j_2})=(2,0)$
  in the above example, the $j_2$ stack has gauge group $U(1)^2$. Further, since $n_2=1  \ {\rm or} \ 2 \bmod3$, so that 
  stack 2 is not degenerate, it is easy to see that either stack $i_1$ or stack $i_2$ is triply degenerate. So in the above example 
  we start with 9 $U(1)$s rather than 6.  Three of the nine $U(1)$ gauge bosons acquire string-scale masses via the generalised
   Green-Schwarz mechanism. These 
include the anomalous $U(1)$s, in particular that associated with baryon number $B=Q_1/3$, and these $U(1)$s survive as global 
symmetries. There remain six gauged $U(1)$ symmetries.  By construction, one is the weak hypercharge $U(1)_Y$, and there is also 
that associated with the sum $X$ of all of the $U(1)$ charges; it is easy to see that all fields are uncharged with respect to $U(1)_X$. 
However, there are four further $U(1)$s, orthogonal to both $U(1)_X$ and $U(1)_Y$ that  also survive. If $n_2=1  \ \bmod3$, for example,
each of the 3 Higgs doublets at the $(2i_1)$ intersection can be used to break spontaneously one $U(1)$, and 
the 6 Higgs doublets at the $(2i_2)$ intersection can be used to break one further $U(1)$; the 4 spontaneously broken $U(1)$s include $U(1)_Y$, 
of course.  Thus, after spontaneous symmetry breaking 
 there is one (unsought) surviving gauged $U(1)$, besides the uncoupled $U(1)_X$,  which does not get a mass 
from either the Green-Schwarz mechanism or the Higgs mechanism. A similar conclusion is reached if $n_2=2  \ \bmod3$. 
The other six-stack models mentioned are even worse in this regard. 
The models with $j_1,k_2$ or $k_1,j_2$ stacks besides $1,2,i_1$ and $i_2$ both have two unwanted extra gauged $U(1)$s after spontaneous symmetry breaking,
 and the $k_1,k_2$ model has no Higgses to break any of the surviving 5 $U(1)$s.
 
 We can do better if we abandon the requirement
  of mass terms for all matter at renormalisable level 
 and consider six-stack models with $i_1$ or $i_2$ stacks, but not both. There are four such models allowed by 
 the constraints. They have stacks $i_1,j_1,j_2,k_2$ or $i_1,j_2,k_1,k_2$ or $i_2,j_1,j_2,k_1$ or $i_2,j_1,k_1,k_2$ besides stacks $1,2$.
The first model has $(n_{j_1}, m_{j_1})= (1,0)$ and $(n_{j_2}, m_{j_2})= (n_2,3)$ in Table 2. If $n_2=1  \bmod3$ then, as above, 
the $i_1$ stack is triply degenerate. There are no other degeneracies, so we start with 8 $U(1)$s. 
In this case 4 get string-scale masses via the generalised Green-Schwarz mechanism, leaving 4 gauged $U(1)$s. 
Since each of the three Higgs doublets at the $(2i_1)$ intersection can spontaneously break one of the $U(1)$s, the only 
 surviving gauged $U(1)$ after spontaneous symmetry breaking is the uncoupled $U(1)_X$. Besides having $n_G=3$ generations of 
 chiral matter (including right-chiral neutrinos), this model has extra vector-like leptonic and quark matter. We find 
 \begin{equation}
3(d^c_L +\bar{d}^c_L)+6(L+\bar{L})+6(e^c_L +\bar{e}^c_L)+3(\nu^c_L +\bar{\nu}^c_L)
\end{equation}
 In addition there are 3 Higgs doublets, 3 charged-singet tachyonic scalars and 3 colour-triplet scalar tachyons. We find that 
 there are mass terms at renormalisable level only for the three generations of $u$ quarks and charged leptons. The gauge 
 coupling constant ratios are identical to those 
 for the six-stack model discussed previously, so it is quite possible that they are consistent with the observed values and a string scale 
 of no more than 3 TeV.
  The  third model has $(n_{j_1}, m_{j_1})= (n_2,3)$ and $(n_{j_2}, m_{j_2})= (2,0)$ in Table 2. The double degeneracy of the $j_2$ stack means 
  that this model has at least one unwanted surviving gauged $U(1)$.  The second and fourth models have no Higgs doublets.

   We now look for models with seven or eight stacks that are free from charged-singlet tachyons. Such models require 
  \begin{equation}
  I_{i_1i_2}=0=I_{j_1j_2}=I_{k_1k_2}
  \end{equation}
for all such intersections. If we assume that there is at most one stack in each class, then the only solutions that are consistent with
eqns (\ref{nj1},\ref{mj1},\ref{nj2}) and (\ref{mj2}), and which possess Higgs doublets,  are those  displayed in Table 3.
\begin{table}
%\begin{eqnarray}
\begin{center}
\begin{tabular}{|c|c|c|c|} \hline \hline
Stack $a$ & $N_a$ & $(n_a,m_a)$ & $\gamma_{\theta,a}$ \\
\hline \hline
1 & 3 & $(1,0)$ & $\alpha ^p {\bf I}_3$ \\
2 & 2 & $(n_2,3)$ & $\alpha ^q {\bf I}_2 $\\
$i_1$ & 1 & $(n_{j}-n_2,-3)$ & $\alpha ^q $\\
$i_2$ & 1 & $(n_{j}-n_2,-3)$ & $\alpha ^q $\\
$j_1$ & 1 & $(n_{j},0)$ & $\alpha ^r $\\
$j_2$ & 1 & $(n_{j},0)$ & $\alpha ^r $\\
$k_1$ & 1 & $(n_{j}-1,0)$ & $\alpha ^p$ \\
$k_2$ & 1 & $(n_{j}-2,0)$ & $\alpha ^p$ \\
\hline \hline
\end{tabular}
\end{center}
\caption{Multiplicities, wrapping numbers and Chan-Paton phases for the D4-brane models with no charged-singlet tachyons. 
($p\neq q \neq r \neq p$).}
%\end{eqnarray} 
\end{table}
Note that we require that $n_j \neq 0$, so that there are $3|n_j|$ Higgs doublet at both the $(2i_1)$ and $(2i_2)$ intersections. 
Note too that the colour-triplet tachyons are also absent in such solutions since $I_{1k_1}=0=I_{1k_2}$. 
It is easy to see that all of these models have 
three generations of chiral matter, including right-chiral neutrinos, and that there is always additional vector-like leptonic, but not quark, matter. 
Precisely how many vector-like leptons are present depends upon the value of the parameter $n_j$, but not on $n_2$:
\begin{eqnarray}
n_j \geq 2  & 6(n_j -1)(e^c_L + \bar{e}^c_L)+6(n_j -1)(\nu^c_L + \bar{\nu}^c_L) +6(2n_j-1)(L + \bar{L}) \\
n_j = 1,0  & 3(e^c_L + \bar{e}^c_L)+3(\nu^c_L + \bar{\nu}^c_L) + 3(2n_j+1)(L + \bar{L}) \\
n_j \leq 0 & 3(1-2n_j)(e^c_L + \bar{e}^c_L)+3(1-2n_j)(\nu^c_L + \bar{\nu}^c_L)+ 3(1-4n_j)(L + \bar{L})
\end{eqnarray}
For any choice of $n_j$ the selection rules discussed earlier permit masses for the three generations of quarks and leptons. 
However, because of the imbalance between the number of $L+\bar{L}$ and the number of $e^c_L + \bar{e}^c_L$ pairs (or the 
number of $\nu^c_L + \bar{\nu}^c_L$ pairs), it is not possible to give renormalisable masses to all vector-like leptonic matter 
 through Yukawa couplings of the type $Le^c_LH, \ \bar{L}\bar{e}^c_LH, \ L\nu^c_LH$ or $\bar{L}\bar{\nu}^c_LH$. The surplus 
 $L + \bar{L}$ pairs can be given masses via nonrenormalisable terms of type $L\bar{L}H_1H_2$; there will be only moderate suppression 
 of such masses by the low string scale. Nonetheless, getting masses for $L+\bar{L}$ pairsabove the electroweak scale is somewhat 
 unnatural.
 
The masses of the tachyonic Higgs doublets \cite{AFIRU2}  do depend on $n_2$ and are given by 
\begin{equation}
m_{H_1}^2=m_{H_2}^2=-\frac{m_s^2}{2\pi}\frac{\epsilon |n_j|(n_2-\delta)}{|\delta||n_j-\delta|}
\label{mH2}
\end{equation}
%where $m_s$ is the string scale;
where $m_{H_{1,2}}$ refers to the doublets at the $(2i_{1,2})$ intersections; $\epsilon$ and $\delta$ are defined in (\ref{eps}) and (\ref{del}).
The above formula for $m_H^2$ is valid so long as $\epsilon \ll 1$, but in any case $m_H^2 \ll m_s^2$ 
is required for consistency of the standard model without major contamination by string effects.
At the string scale, the value  of the gauge coupling constant $\alpha_a(m_s)$ for the gauge bosons asociated with stack $a$ is inversely proportional
to the length $l_a$ of the 1-cycle wrapped by the stack. For $\epsilon \ll 1$ 
\begin{equation}
l_a \simeq 2\pi |n_aR_1 -m_aR_2|
\end{equation}
 Ratios of the gauge coupling constants are independent of the Type II string coupling constant $\lambda_{II}$. Thus
\begin{equation}
\frac{\alpha_3(m_s)}{\alpha_2(m_s)}=|n_2 -3R_2/R_1| = |\delta|
\label{32}
\end{equation}
Also, since 
\begin{equation}
\frac{1}{\alpha_Y} = \frac{1}{3\alpha_3}+\frac{1}{2\alpha_2}+\frac{1}{\alpha_{i_1}}+\frac{1}{\alpha_{j_1}}+\frac{1}{\alpha_{k_1}}
\end{equation}
we have that
\begin{equation} 
\frac{\alpha_3(m_s)}{\alpha_Y(m_s)}=\frac{1}{3}+\frac{1}{2}|\delta|+|n_j-\delta|+ |n_j| + |n_j -1|
\label{3Y}
\end{equation}
Consistency with a low string scale $m_s$ requires that these ratios do not differ greatly from the values measured \cite{cernyellow} at the 
electroweak scale $m_Z$
\begin{eqnarray}
\frac{\alpha_3(m_Z)}{\alpha_2(m_Z)} = 3.54 \\
\frac{\alpha_3(m_Z)}{\alpha_Y(m_Z)} = 11.8 \label{3Yexp}
\end{eqnarray}
It is easy to see that when $\delta =-3.54$ the value $n_j= -4$ in (\ref{3Y}) comes very close to the value (\ref{3Yexp}), and
even $n_j= 2$ comes fairly close. Interestingly, in the case $n_j= 2$ the model has only seven stacks. 
When $\delta =+3.54$ rough agreement occurs for $n_j= 5,-2$. 
The case $n_j= 1$ gives the value for the six-stack models derived above;
% and, as 
%already observed, neither value of $\delta =\pm3.54$ gives a value of $\alpha_3/\alpha_Y$ that is close to the measured value (\ref{3Yexp}).
$\delta = -3.54$ gives a value  $\alpha_3/\alpha_Y=7.64$ that is closer to the measured value (\ref{3Yexp}) than that given by $\delta = +3.54$.

In conclusion, we find that requiring that the gauge boson  associated with weak hypercharge does {\it not} acquire 
a string-scale mass leads to a unique, one-parameter $(n_2)$ family of six-stack, intersecting D4-brane models, if we are to get three chiral generations
of matter all of which are coupled to the tachyonic Higgs bosons that generate masses when the electroweak symmetry is spontaneously broken.
Such models all have extra vector-like leptons, as well as charged-singlet scalar tachyons.
 They also have at least one surviving (unwanted) coupled, gauged $U(1)$ symmetry after spontaneous symmetry breaking. 
Relaxing the requirement that all matter has the Higgs couplings necessary to generate masses at renormalisable level
 allows us to construct models without unwanted $U(1)$ gauge symmetries. In this case there are also 
  colour-triplet scalar tachyons as well as vector-like $d$ quark matter.  
Demanding both that there are mass terms at renormalisable level for all matter and that there are no charged-singlet tachyons 
generally requires models with at least eight stacks of intersecting branes, and these have one extra parameter $(n_j)$. When $n_j=-4$ the standard model
 gauge coupling ratios have values at the string scale that are very close to those measured at the electroweak scale. In this case, therefore,
  the string scale 
 cannot be far from the electroweak scale, and  stringy signatures should appear below this scale. Specifically,
  towers of
  vector-like  fermionic and bosonic matter, with mass scale set by the Higgs bosons, are unavoidable in all intersecting brane models \cite{AFIRU2}. 
  Also, D4-brane models such as ours, in which the orbifold point group acts supersymmetrically on the transverse space $(T^2 \times T^2)/\IZ_3$, 
  have ${\mathcal N}=2$ supersymmetric gauge supermultiplets. However, since the spectrum is not supersymmetric, we expect masses to arise from loop 
  effects involving the non-gauge matter. Again there is every reason to expect these particles to have masses below the string scale.
  Three of the 
  %All bar two of the 
  original $U(1)$ gauge bosons, including those of the anomalous $U(1)$s,  acquire string-scale masses via the generalised Green-Schwarz mechanism.
  % masses\footnote{The two exceptions are 
  %the weak hypercharge $U(1)_Y$, by construction, and also that associated with the sum of all of the $U(1)$ charges; it is easy to see that  
  %all fields are uncharged with respect to this $U(1)$.}, and survive as global symmetries of the four-dimensional theory. 
  In particular, baryon 
  number $B=Q_1/3$ is anomalous and survives as a global symmetry. Consequently, the proton is stable despite the low string scale. 
  As before \cite{BKL1},
  the Higgs boson fields are also charged under some of the anomalous $U(1)$s that survive as global symmetries. Thus a keV-scale axion is unavoidable. 
  It is expected that TeV-scale $Z'$ vector bosons will be observable at future colliders, and precision electroweak data (on the $\rho$-parameter) 
  already constrain \cite{GIIQ} the string scale  to be at least 1.5 TeV. 
  However, there are generally ($2|n_j|+|n_j -1|+|n_j -2|-2 \geq 1$) surviving, unwanted $U(1)$ gauge symmetries after spontaneous 
  symmetry breaking.

\section*{Acknowledgements}
This research is supported in part by PPARC and the German-Israeli 
Foundation for Scientific Research (GIF).


\newpage

\begin{thebibliography}{99}
\bibitem{UTCA} G. Aldazabal, L. E. Ib\'{a}\~{n}ez and F. Quevedo, JHEP 0001(2000)031, hep-th/9909172,
JHEP 0002 (2000) 015, hep-th/0005067; 
I. Antoniadis, E. Kiritsis and T. Tomaras, Phys. Lett. B486 (2000) 186, hep-th/0004214;
 D. Bailin, G.V. Kraniotis, A. Love, Phys. Lett. B502 (2001) 209, hep-th/0011289;
C. Bachas, hep-th/9503030;   R. Blumenhagen, L. G\"{o}rlich, B. K\"{o}rs and 
D. L\"{u}st, JHEP 0010 (2000) 006, hep-th/0007024;
Z. Kakushadze, Phys. Rev. D59 (1999) 045007;
 Z. Kakushadze, G. Shiu and S.-H. Henry Tye, Nucl. Phys. B533(1998)25;
J. Lykken, E. Poppitz, S. P. Trivedi, Nucl. Phys. B543 (1999) 105; 
I. Antoniadis, E. Dudas, A. Sagnotti, Phys. Lett.B 464 (1999) 38; S. Sugimoto, 
Prog. Theor. Phys. 102 (1999) 685; C. Angelantonj, Nucl. Phys.B 566 (2000) 126;
\bibitem{BDL}M. Berkooz, M. R. Douglas and R. G. Leigh, Nuclear Physics B480 (1996) 265, hep-th/9606139
\bibitem{ADD}N. Arkani-Hamed, S. Dimopoulos and G.R. Dvali, Phys. Lett. B429 (1998) 263, hep-ph/9803315; 
I. Antoniadis, N. Arkani-Hamed, S. Dimopoulos and G.R.Dvali, Phys. Lett.B 436 (1998) 257, hep-ph/9804398.
\bibitem{BKLO}R. Blumenhagen, B. K\"{o}rs, D. L\"{u}st and T. Ott, Nucl. Phys.B616 (2001) 3, hep-th/0107138 
\bibitem{Cvetic}M. Cveti\v{c}, P. Langacker and A. M. Uranga, Nucl. Phys. B615 (2001) 3, hep-th/0107166;
M. Cveti\v{c}, P. Langacker and G. Shiu, hep-ph/0205252, hep-th/0206115
\bibitem{IMR}L. E. Ib\'{a}\~{n}ez, F. Marchesano and R. Rabadan, JHEP 0111 (2001) 002, hep-th/0105155
\bibitem{CIM}D. Cremades, L. E. Ib\'{a}\~{n}ez and F. Marchesano, hep-th/0205074
\bibitem{IBM}R. Blumenhagen, V. Braun , B. K\"{o}rs and D. L\"{u}st, JHEP 0207 (2002) 026, hep-th/0206038 ;
G. Aldazabal, L. E. Ib\'{a}\~{n}ez, A.M. Uranga, hep-ph/0205250;
G. Honecker, JHEP 0201 (2002) 025, hep-th/0201037, hep-th/0112174;
H. Kataoka and M. Shimojo, hep-th/0112247; 
C. Kokorelis, hep-th/0203187,  hep-th/0205147;
J. R. Ellis, P. Kanti, D.V. Nanopoulos, hep-th/0206087 
\bibitem{BKL1}D. Bailin, G. V. Kraniotis and A. Love, Physics Letters B530 (2002) 202, hep-th/0108131
\bibitem{IRU}L. E. Ib\'{a}\~{n}ez, R. Rabad\'{a}n and A. M. Uranga, Nucl. Phys. B542 (1999) 112, hep-th/9808139
\bibitem{AFIRU2} G. Aldazabal, S. Franco, L. E. Ib\'{a}\~{n}ez, R. Rabad\'{a}n and A. M. Uranga, JHEP 0102 (2001) 047, hep-ph/0011132
\bibitem{AFIRU1} G. Aldazabal, S. Franco, L. E. Ib\'{a}\~{n}ez, R. Rabad\'{a}n and A. M. Uranga, hep-th/0001083
\bibitem{GN}D. M. Ghilencia and H. P. Nilles, hep-th/0204261
\bibitem{cernyellow}Reports of the working groups on precision calculations for LEP2 Physics,
S. Jadach, G. Passarino, R. Pittau (Eds.), CERN Yellow Report, CERN, 2000-009
\bibitem{GIIQ} D. M. Ghilencia, L. E. Ib\'{a}\~{n}ez, N. Irges and F. Quevedo, hep-ph/0205083

\end{thebibliography}
\end{document}